**Reframing the Free Will Debate: The Universe is Not Deterministic.**

Henry D. Potter[1], George F.R. Ellis[2] & Kevin J. Mitchell[1]

[1]Smurfit Institute of Genetics and Institute of Neuroscience, Trinity College Dublin, Dublin 2, Ireland.

[2]Mathematics Department, University of Cape Town, Rondebosch, Cape Town 7701, South Africa.

Henry Potter: https://orcid.org/0000-0003-0765-5205
Kevin Mitchell: https://orcid.org/0000-0002-3433-4530

Corresponding author: kevin.mitchell@tcd.ie

**Acknowledgements:** HP was supported by a Provost's Fund grant (1481.9050961) from Trinity College Dublin.

**Conflicts of interest:** The authors declare no conflicts of interest.

*"It ain't what you don't know that gets you into trouble.*
*It's what you know for sure, that just ain't so"  (Mark Twain)*

**Abstract**

Free will discourse is primarily centred around the thesis of determinism. Much of the literature takes determinism as its starting premise, assuming it true "for the sake of discussion", and then proceeds to present arguments for why, if determinism *is* true, free will would be either possible or impossible. This is reflected in the theoretical terrain of the debate, with the primary distinction currently being between compatibilists and incompatibilists and not, as one might expect, between free will realists and skeptics. The aim of this paper is twofold. First, we argue that there is no reason to accept such a framing. We show that, on the basis of modern physics, there is no good evidence that physical determinism (of any variety) provides an accurate description of our universe and lots of evidence against such a view. Moreover, we show that this analysis extends equally to the sort of 'indeterministic' worldviews endorsed by many libertarian philosophers (and their skeptics) – a worldview which we refer to as determinism-plus-randomness. The paper's secondary aim is therefore to present an alternative conception of indeterminism, which is more in line with the empirical evidence from physics. It is this indeterministic worldview, we suggest, that ought to be the central focus of a reframed philosophy of free will.

## 1. Introduction

The thesis of determinism – "the thesis that there is at any instant exactly one physically possible future" (van Inwagen, 1983:3) – plays a central, and often defining, role in the philosophy of free will. This is evident from the debate's current theoretical landscape. As Müller et al. (2019:6) write:

> "The free will debate standardly has its focus on the consequences of determinism for free agency. This is obvious when we look at the major distinction in the free will debate. One might expect that that distinction concerns theories that affirm freedom versus those that deny it. However, the major distinction among free will theorists is whether they are *compatibilists* or *incompatibilists* with regard to freedom and determinism. Thus, the central question that splits the debate into two opposing parties is about the implications of determinism for freedom"

Or, as philosopher Peter van Inwagen (1983:55) puts it: "The main contested question in current discussions of free will is not, as one might expect, whether we *have* free will. It is whether free will is compatible with determinism."

In part, the reason why the theoretical terrain takes this form, and why free will discourse focuses so extensively on determinism, is historical. Some of the earliest known articulations in Western thought of the very idea that there might, in fact, be a *problem* of free will to contend with at all – that is, that there may be any reason to doubt the veridicality of our intuitive feeling of having free will – came within the context of various early forms of determinism. The early Greek materialist philosophers, Leucippus and Democritus, for example, formulated an influential atomistic worldview in which everything, including humans, was said to be ultimately made up of 'atoms' moving through 'the void' along necessitated pathways. The upshot, as Leucippus stated, was that 'nothing occurs at random, but everything for a reason and by necessity' (Edmunds, 1972). In other words, everything that happens was physically (and logically) determined to happen by what went before it. Later, Stoic philosophers added to this worldview the concept of laws of Nature (or laws of God), which they took to govern the trajectory of these atoms in the void, thereby providing an explanation for their deterministic nature (O'Connor & Franklin, 2022).

Philosophers of the time – including the Stoics themselves, as well as Aristotle and the Epicureans – recognised that such a deterministic worldview posed at least a *prima facie* threat to our intuitive free will. The Stoics, led most notably by the philosopher Chrysippus, sought to resolve the tension by arguing that what is needed for our actions to be 'up to us', in the sense required for free will, is just that their causes 'flow through us' – and hence our free will would not in fact be negated by determinism (a forerunner to the position now known as compatibilism). Aristotle and Epicurus, on the other hand, insisted that truly free actions could not be physically necessitated in this manner, and so argued that some form of indeterminism *must* exist in the universe if our actions are to be free (a forerunner to the position now known as incompatibilism) (O'Connor & Franklin, 2022). Whichever of these options one favours, the point for present purposes is that these early formulations of the problem of free will clearly took it to be *the problem of free will and determinism* – thereby laying the foundations for the conceptual terrain we see today.

In the intervening millennia, however, the discovery of quantum mechanics has led many philosophers, and most physicists (Schlosshauer et al., 2013), to officially reject the truth of determinism; few contemporary philosophers would now publicly subscribe to a thesis of complete physical determinism. Yet it continues to play a central role in our philosophy of free will. Why is this? Why do we still dedicate the vast majority of our attention to whether free will is compatible with determinism, when most would outwardly reject determinism itself? There are perhaps a few reasons.

First, a small number of philosophers explicitly state that their focus on determinism stems primarily from a desire to safeguard their intuitions about free will, and – often more explicitly – moral responsibility, from certain potential future outcomes of scientific investigation. As John Martin Fischer and Mark Ravizza (1998:15) write:

> "Our contention is that … even if we discovered that causal determinism were true, there is a strong tendency to think that this sort of discovery should not make us abandon our view of ourselves as persons and morally responsible agents."

On the contrary, Fischer later writes:

> "My basic views of myself and others as free and responsible are and *should be* resilient with respect to such a discovery about the arcane and "close" facts pertaining to the generalizations of physics." (Fischer, 2007:45, *our emphasis*)

Free will philosophers of this persuasion might therefore continue to at least entertain the possibility of determinism for the purposes of their research, agreeing that such a thesis looks highly unlikely to be true of *our* world, but motivated by the belief that "[i]t is notoriously difficult to predict how future science will turn out… [so] it might be useful to have an answer to the question [of how we might retain our view of ourselves as free and responsible given determinism] in advance of the scientific issues getting sorted out" (Fischer et al., 2007:2).

Most philosophers take a different tack, however. They justify their focus on determinism on empirical grounds by appealing to some variant of the claim that, though strict *universal* determinism (i.e., a single, inevitable and predetermined timeline of all events in the universe for all time) might not be true, something close enough *is* deemed by physicists to be true. Indeed, Fischer and Ravizza (1998:15) *also* make appeal to this line of reasoning themselves, stating that:

> "There is an additional reason to focus our attention on causal determinism.[1] Although contemporary physicists tend to believe that causal determinism is false, they believe that something very much like it is true: a doctrine we shall call "almost causal determinism." On this view, macroscopic events are not, strictly speaking, causally determined, but they are very close to being determined"

In this way, the thesis of determinism persists in the contemporary free will debate, rarely in its strongest form (which we will call 'universal determinism'), but more commonly in some weaker form (which we will call 'near-determinism' – although it has been variously referred to in the literature to as 'almost causal determinism' (*ibid*), 'neural-level determinism' (Pereboom, 2022:8), 'macro determinism' (Honderich, 2001:465), 'hard-enough determinism' (Caruso, 2012:4), and 'for all practical purposes' determinism (Kane, 2001:8)).

Near-determinism itself tends to take one of two forms. The first is the thesis of classical determinism. This is the view that indeterminism does indeed exist at quantum scales of reality, but that it is of no relevance whatsoever to the problem of free will; either, because the indeterminism of the quantum world completely disappears as subatomic particles combine to form the everyday objects of the macroscopic (or "classical") world, or because any quantum effects that *do* survive to the classical limit are so small and rare as to be entirely irrelevant to the processes of neural decision-making and action. As philosopher

[1] Note that there is an important distinction, not obvious from this quote, between causal determinism and physical pre-determinism or universal determinism (See Section 6.4).

Derk Pereboom (2022:8) puts it, classical determinism (or, for him, "neural-level determinism") is the view that "quantum micro-indeterminacies… are ordered with enough redundancy so that at the neural level, indeterminacy all but vanishes".

In other words, the thesis of classical determinism is the view that, regardless of what is going on in quantum mechanics, the macroscopic world of brains, bodies and behaviour still evolves deterministically. It states that for objects and systems at this level of reality, which includes the putative subjects of free will, it remains the case "that there is at any instant exactly one physically possible future" (van Inwagen, 1983:3).

There is also a second, weaker, version of near-determinism at play in the free will literature, which we will call 'determinism-plus-randomness'. On this view, quantum indeterminacies are *not* seen as irrelevant to goings-on at the neural level. It is *not* taken to be the case that "indeterminacy all but vanishes" at macroscopic scales of reality. On the contrary, it is assumed that quantum effects may occasionally get amplified within the brain, introducing a non-negligible element of randomness into the macroscopic chains of causation involved in decision-making and action, and thereby rendering the subsequent decision and action outcomes causally *undetermined*.

Pereboom (2022:8) articulates this perspective nicely when he describes how:

> "[f]or alternative possibilities [in decision-making] to be significantly probable, there would have to be mechanisms that facilitate the "percolating up" of significant microlevel indeterminacies to the neural level, on the analogy of a Geiger counter that senses microlevel events and registers them at the level of the moving of a macrolevel indicator."

This 'percolating up' model is perhaps the dominant way that indeterminism is conceptualised in the free will literature, common even among libertarian philosophers who seek to reject the thesis of determinism. And yet it is an implicitly near-deterministic worldview. By endorsing a picture in which quantum indeterminacies occasionally 'percolate up' to disrupt the deterministic processes unfolding at macroscopic scales, one is tacitly accepting that, for the most part (e.g., on local timescales), determinism holds true in the brain. That is, one is accepting determinism as "Nature's default mode" (Earman, 2008:817) at the level of neural activity, even in spite of the occasional possibility for disruption 'from below'.

Such a view is clearly evident in McKenna and Pereboom's (2016:16) description of how an "indeterministic" world (or, what we are proposing to call, a 'determinism-plus-randomness' world) is supposed to differ from a (universally or classically) deterministic one:

> *Consider the following model of a deterministic world, Wd, followed by a model of an indeterministic world, Wi. Let "e" represent an event. Let "—" represent a causally* *deterministic (d) relation between events. And let "…" represent an indeterministic (i) relation between events. Now consider each world:*

*(Wd): e1—e2—e3—e4—e5—e6—e7—e8, and so on with only d relations* (1)

*(Wi): e1—e2 … e3—e4—e5—e6—e7—e8, and so on with only d relations* (2)

On this model, indeterminism is conceptualised as something that gets intermittently *added* into an *otherwise deterministic* causal chain of events. It is for this reason that we refer to this popular worldview as 'determinism-plus-randomness', and treat it as a species of near-determinism, rather than viewing it as a *rejection* of determinism itself (as many of its proponents do).

These three metaphysical perspectives – universal determinism, classical determinism, and determinism-plus-randomness – collectively dominate the range of worldviews currently under consideration in the (naturalistic) free will literature. Here, we argue that none of them is supported by the empirical evidence.

In so doing, we hope to encourage a fundamental re-framing of the free will debate toward something that is more aligned with modern physics in two key respects: first, in its comprehensive rejection of physical determinism; and, second, in the way that indeterminism gets conceptualised within the debate. Importantly for free will philosophy, we aim to show that the empirical evidence supports a picture of indeterminacy that does not merely present as random *additions* to an otherwise deterministic universe, but rather as a pervasive indefiniteness in the current states and in the future evolution of physical systems. What this means for the putative subjects of free will is that the future, as a whole, is simply *underdetermined by its current state*; alternative possible futures, from the perspective of physics, are both inevitable and innumerable.

We end by proposing that, under such a worldview, the important question for free will philosophy is no longer 'where does the freedom come from?' but rather 'where does the control come from?' The appropriate focus is not on whether an agent 'could have done otherwise' than they did; but, rather, on how they managed to prevent all of the other physically possible 'otherwises' from happening, such that they were able to do what they wanted to do or what they *chose* to do. Reframing the free will debate in this way opens up a series of new problems, new questions, and a new way of conceptualising this age-old debate, which, we hope, can support future progress on this vexed issue.

In Section 2, we lay out the evidence and arguments against universal determinism, in particular by showing that the processes governing the evolution of quantum systems are genuinely indeterministic. In Section 3, we turn our attention to classical determinism. We consider a number of arguments claiming that, despite the existence of fundamental quantum indeterminacies, macroscopic (or 'classical') systems nevertheless evolve according to deterministic laws and processes. We show that none of these arguments is well supported and that the idea that classical physics is deterministic is not in fact a *result* of physics, but merely a convenient idealisation. Moreover, we present additional evidence that positively argues *against* classical determinism, and conclude that not only is there no good reason for philosophers to start with this premise, there are strong reasons not to. In Section 4, we consider the view of indeterminacy – "determinism-plus-randomness" – which is often taken as the starting point for libertarian arguments. This view takes deterministic evolution of physical systems as the default, but allows that it is occasionally interrupted by isolated random or undetermined events. This poses some well-known problems for libertarian views, as it is not easy to see how agents can be said to be in control of any action or choice that results from such an undetermined process, where what happens is ultimately just settled by the outcome of random microscopic events within them. Based on the evidence surveyed in Section 3, we present an alternative view of indeterminacy, which we refer to as "pervasive indefiniteness". Under this view, the future states of certain complex systems, such as ourselves, are *under-determined* by their current state and microscopic laws. In Section 5, we consider the implications of this physical view for the free will debate, arguing that it renders much of the debate's traditional positions and dividing lines effectively moot, and opens the door to a kind of libertarianism that can actually explain the evolution of agents. Lastly, in Section 6, we examine and respond to a number of possible objections to our arguments.

## 2. The argument against universal determinism

Universal determinism is the strongest version of the deterministic thesis. It states that "there is at any instant exactly one physically possible future" (van Inwagen, 1983:3). This is taken to be the case because the future evolution of the universe is understood to be physically necessitated by its current state and the laws of nature. Or, in slightly different terms, it is "the view that, for any given time *t*, a complete statement of the facts at *t*, together with a complete statement of the laws of nature, entails every truth as to what happens after *t*" (Palmer, 2014:4; also Fischer & Ravizza, 1998:14).[2]

This was the original version of the thesis, implied by the atomistic worldview of the early Greek materialists and the Stoic philosophers. It is also the version of the thesis articulated in the influential and evocative metaphor of Laplace's demon, an omniscient being for whom the state of the universe across all moments of time would be laid bare at once. Laplace describes the demon as:

> *An intelligence which, for one given instant, would know all the forces by which nature is animated and the respective situation of the entities which compose it, [and] if besides it were sufficiently vast to submit all these data to mathematical analysis, would encompass in the same formula the movements of the largest bodies in the universe and those of the lightest atom; for it, nothing would be uncertain and the*

[2] Note that this could only apply to a *closed* system. For the purposes of argument, we will assume (as proponents of determinism do) that the entire universe can be viewed as such a system. However, it should be noted that some models of spacetime, such as anti-de Sitter space (Hawking & Ellis 2024), are not single closed systems.

> *future, as the past, would be present to its eyes.* (Laplace, 1814:3-4, translation from van Strien, 2014).

Some philosophers still accept the notion of universal determinism implied by the conceivability of such a demon, and many others are at least willing to entertain it as the premise for arguments about free will and moral responsibility. For example, though they differ on its implications, Gregg Caruso and Daniel Dennett, in a 2021 book, start out by both accepting the premise that: “facts about the remote past in conjunction with laws of nature entail that there is only one unique future” (p.5). (Again, taking the entire universe as a single, closed system; the future states of more local, open systems may of course always be affected by information entering from outside the system).

We argue here that this premise is not sustainable; the universe is not deterministic in that fashion, either at quantum or classical levels. We start with the quantum level.

### 2.1 Quantum physics is not deterministic

The universal determinism worldview entails, at least in principle, “the claim that all processes can be fully described through a set of fundamental laws of nature [typically modeled using sets of differential equations], which always have a unique solution for given initial conditions” (van Strien, 2021:2). The thesis therefore relies on two key assumptions – both of which were undermined in the twentieth century by the discovery of fundamental indeterminacy in quantum mechanics.

Universal determinism’s first critical assumption is that systems (including whole universes) exist in precisely defined (or hypothetically *definable*) states at every given instant of time. If they did not, then we would not be able to specify the initial conditions of the system from which it uniquely (i.e., deterministically) evolves. That is, there would be no physical reality that corresponds to the claim that Laplace’s Demon could “know all the forces by which nature is animated and the respective situation of the entities which compose it”.

This assumption is straightforwardly ruled out by the Heisenberg Uncertainty Principle (HUP). The HUP describes the fact that in a system of elements with wave-like properties (as is the case in quantum systems) the operators corresponding to various physical observables, most famously including the position and momentum of particles, are related to each other by a Fourier transformation. As a result, the more precisely one of these “conjugate variables” is defined, the less definition the other variable has. In other words, the HUP tells us that it is physically impossible to give ‘a complete statement of the facts at *t*’ for the entire universe – as is a stated requirement of universal determinism (Fischer & Ravizza, 1998) – because ‘completeness’ in one area of the system *necessarily* comes at the cost of ‘completeness’ in another.

It should be noted that the use of the word “uncertainty” to describe this principle can be misleading as it may give the impression of a purely epistemic phenomenon. That is, it may give the impression that the Uncertainty Principle refers to the inability of an observer to precisely *measure* both variables at once. However, the principle (originally dubbed the Indeterminacy Principle by Heisenberg) is emphatically ontic in nature. Conjugate variables simply cannot simultaneously exist with infinite precision, meaning it is physically impossible for a system – or the universe as a whole – to exist in an infinitely precise and exhaustively definable state at any given time.

This is demonstrated in a concrete way by the observation of "zero point energy". When the temperature of a system is reduced to absolute zero (the absolute lowest it can go), one might reasonably assume, given that the motion of particles is what generates temperature in the first place, that the motion of the system's constituent particles would come to a standstill and that all energy in the system would be lost. This is not the case, however. "Zero point energy" is the observation that there always remains some energy in the system, even at absolute zero (Milonni, 1994). The explanation for this is that, if it were not the case, then the conjoint precision of the momentum and the position of particles (as possible manifestations of the pervading quantum fields) would be infinite, and that would violate the HUP. Instead, then, energy remains in the system due to the irreducible probability of quantum fluctuations occurring as a result of the HUP. The empirical evidence for zero point energy is thus consistent with the interpretation that the HUP is an ontic – and not merely epistemic – feature of our world.

The second key assumption of universal determinism is that any isolated physical system – and, indeed, the entire universe – evolves *necessarily* from one timepoint to the next. Mathematically speaking, it is the assumption that the transitions between 'states' of the system are derived from (or described by) sets of differential equations which always have a unique solution for the given antecedent state. This is the notion of a deterministic law of nature, which is what ultimately underlies the universal determinist's claim that *if* Laplace's Demon had access to the position and momenta of every particle in the universe, it could then "submit all these data to mathematical analysis" and deduce "the [future] movements of the largest bodies in the universe and those of the lightest atom". The relevant assumption here being that such a 'mathematical analysis' would necessarily yield a *unique* solution (and, also, that such a mathematical formalism fully captures how the universe itself reaches the next state).

Again, this assumption is challenged by research in quantum physics. The evolution of quantum systems from one timepoint to the next – i.e., the transition between 'states' – is fundamentally indeterminate, not only as a feature of the initial indefiniteness described by the HUP, but more generally, in that the state of a system is described by a *probability density function*, rather than with fixed, precise values for every parameter. The Schrödinger equation is what provides us with the means to calculate how such a quantum system will evolve through time, and it is often described as being deterministic. This is true insofar as *the equation itself* does not admit any randomness and theoretically has a unique (unitary) solution for any future time-point. However, the solution it gives is still a distribution of probabilities, not a single actuality. What the equation tells us is how the *probabilities* of observing the system in one state or another will evolve over time, if the system is not observed. When we want to see how the physical system *actually* evolves, what we get in a single trial is a "collapse" to one of those possible states – a collapse that seems to be genuinely un(der-)determined by any antecedent conditions.[3] (See Section 6 on Possible Objections for comments on unitarity and the conservation of information).

There is no reason to think – as is sometimes suggested – that the resolution of this indefiniteness in the system relies on *an observer*. It arises any time some physical interactions force the system into a definite state. Empirically, across many such trials, the Schrödinger equation very accurately predicts the frequency of different outcomes (for simple systems at least). But in any given trial, the outcome seems to be genuinely probabilistic – a random draw from the probability distribution. In that sense, how the potentialities of a quantum system become the actualities that obtain and that we observe, in practice, does seem to be truly not fully determined in advance.

A variety of theories have been proposed to try to rescue determinism within such systems. These include ones that invoke hidden variables of one kind or another, which are taken as actually explaining (in virtue of causally determining) which specific outcome arises in any specific instance (reviewed in Earman, 2004). However, no evidence for such hidden variables has ever been found, and their existence seems to be ruled out by Bell's theorem (under locality) (Bell, 1964), which has been empirically supported time and again (e.g., Acín & Masanes, 2016; Abellán et al., 2018).

An alternative approach – dubbed the Many Worlds Hypothesis – is to simply assume that every possible value in the probability density function actually gets realised, in a process that generates separate universes for each possibility at each such event (Everett, 1957). Such a scheme can be made mathematically consistent and is even parsimonious in the sense that it assumes the Schrödinger equation is all you need to capture the evolution of a quantum system, with no mysterious "collapse" when things happen. However, there seems little reason to take it seriously as a physical reality or even as a metaphysical possibility. Physically, it assumes that the entire universe can be described by a single Schrödinger equation, but there are strong arguments against this notion (Drossel, 2017; Ellis 2023). Metaphysically, it trades the problematic notion of collapse for the extravagance of infinite universes, which, like a girlfriend who goes to another school, remain unknowable to us. And more practically, this hypothesis simply does no work in explaining *which possibility occurs in the universe we are experiencing.* As proponent of the Many Worlds Hypothesis, Sean Carroll puts it, under this interpretation: "the underlying laws are perfectly deterministic, but what happens along any specific history is irreducibly probabilistic".[4]

We take it that it is therefore generally accepted that there exists fundamental indeterminacy in quantum systems and that the evolution of quantum systems is essentially non-deterministic.[5] The upshot, as many philosophers of free will seem to agree, is that the thesis of universal determinism is highly implausible as a description of our universe.[6]

[3] Note that we make no commitment here to the ontological status of either the wave function or its "collapse". We use the term, here and later, to refer to the resolution of a probability distribution to a single actuality.
[4] https://www.preposterousuniverse.com/blog/2011/12/05/on-determinism/

### 2.2 New possibilities emerge as the universe expands

Before moving on to the more contested thesis of classical determinism, it is worth briefly pausing to mention the status of a claim that is often seen in the free will literature and that seems to go hand-in-hand with universal determinism. This is the claim that the conditions of the early universe, in the moments after the Big Bang, causally fixed everything that has happened and will happen in our universe. Or, in an alternative framing, it is the claim that, in some sense, the Last Scattering Surface of the early universe could have contained all of the information necessary for a hypothetical, Laplacean being to predict every future event and state – including the Cretaceous extinction event, the invention of iPhones, and every single choice and action an individual human will take in their lifetime. Here is philosopher Christian List describing (though not endorsing) such a claim:

> *A second kind of argument derives the unreality of free will from the claim that the laws of physics may be deterministic, meaning that the initial state of the universe, say at the Big Bang, pre-determined all subsequent events; so, there would be no room for alternative possibilities to choose from. You may think that you had a choice whether to read this article or not, but in reality, your decision was made for you by the world's initial conditions.* (Caruso et al., 2020:2)

Thankfully this unsettling proposal has no actual basis in physical reality. First, the indefiniteness of quantum systems described above renders it both impossible *and* unnecessary, as explained by Georges Lemaître, who first proposed the idea of the universe beginning from a singular quantum state:

> "*Clearly the initial quantum could not conceal in itself the whole course of evolution; but, according to the principle of indeterminacy, that is not necessary. Our world is now understood to be a world where something really happens; the whole story of the world need not have been written down in the first quantum like a song on the disc of a phonograph. The whole matter of the world must have been present at the beginning, but the story it has to tell may be written step by step.*" (1931:706).

Second, Lemaître's view can be supported by an informational perspective. Because information requires a physical substrate (Landauer, 1991), the amount of information that can be held in any finite region of space is limited. In particular, it is limited by the Bekenstein bound, which caps the peak entropy of a system of finite size, and thus its capacity to hold information (Shannon & Weaver, 1949; Bekenstein, 2004). In physical terms, the information storage is fundamentally limited by energy and size (specifically surface area). Increasing the energy in a finite region of space beyond a certain point would result in the system collapsing into a black hole, with the surface of the black hole saturating the Bekenstein bound. For a hypothetical Laplacean demon to be able to predict the invention of iPhones from the Last Scattering Surface of the early universe, however, this finite region of space would need to have contained an effectively infinite amount of information, thus saturating the Bekenstein bound and resulting in a black hole. This provides strong evidence that such a proposal is indeed physically impossible.

[5] This would also mean that these systems are not time-symmetric either, meaning that information *really is* created and destroyed as time proceeds. We address this point further in the Possible Objections section below.
[6] Of course, how exactly to interpret the fundamental indeterminacy of quantum systems is far from settled. Not only are there attempts to interpret it deterministically (which we have suggested are unsuccessful), but different indeterministic interpretations also exist. Recent work by Jacob Barandes (2023; 2025), for example, demonstrates how one can fundamentally reformulate quantum theory "in the language of trajectories unfolding stochastically in configuration spaces", in a way that does not rely on superpositions or wave function collapses, while recovering the full range of quantum phenomena.

Lastly, claims about the early universe pre-determining everything that we do are *also* undermined by the fact that the universe has continuously expanded since the Big Bang, creating an ever-expanding possibility space of states that it could be in (Figure 1). This rate of expansion outstrips the rate of equilibration, meaning that while entropy continues to increase, so too does information (Layzer, 1975, 2021). As Arthur Eddington said, in 1935: "*The expansion of the universe creates new possibilities of distribution faster than the atoms can work through them*." (p.66) It is therefore not plausible that the outcomes of all future physical events could be predetermined, *fixed* by, or encoded *in* the state of the universe at one point in time when the possibility space in which these futures will evolve does not yet even exist. As we will argue later on, this point applies not only to the origins of the universe, but also to the state of the universe at any subsequent timepoint (including the present).

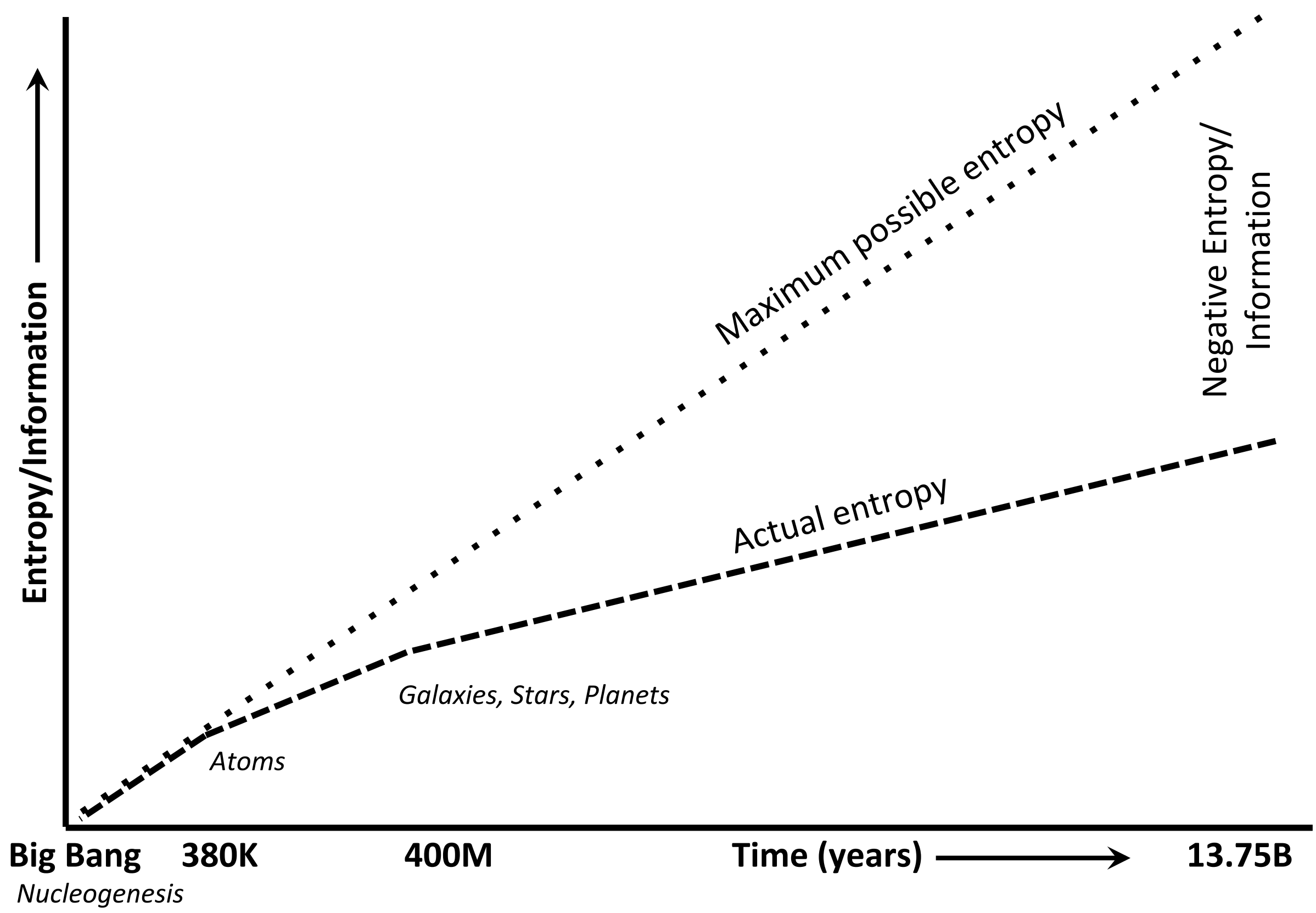

**Figure 1**. At the Big Bang, the universe was in a state of very low entropy, relative to later times. As it expanded, and atoms, then galaxies, stars, and planets emerged, the processes of physical equilibration lagged behind the pace of expansion of physical space (and possibility space). This means that even as entropy increased, so too did information. (Based on a conceptual illustration by Robert Doyle, depicting the scenario described by David Layzer in his 1975 article, *The Arrow of Time*).

Most philosophers of free will seem to accept that the extremely strong evidence for fundamental indeterminacy at quantum levels ought to put the nail in the coffin of universal determinism. What this means for systems at the so-called "classical" level, where everyday macroscopic objects reside, is more contested, however. It is to this we now turn.

## 3. The case against classical determinism

Intuitively, it might seem obvious that, if determinism does not hold at the quantum level, then it cannot hold at the classical level, either. At least not in the strict sense in which, for macroscopic objects and systems, "there is at any instant exactly one physically possible future", if that future is taken to be defined with microscopic precision (van Inwagen, 1983:3). After all, everything is made up of atoms and subatomic particles and, ultimately, quantum fields. Quantum indeterminacy should thus be constantly at play in the deepest reality of every system, including those systems that may be the putative subjects of free will.

Yet, the relevance of quantum indeterminacy for goings-on in the macroscopic world is often dismissed by philosophers of free will (and by many physicists, too). These arguments lean on a number of different ideas, but they all lead to the conclusion that classical levels are somehow *insulated* from the indeterminate goings-on at quantum levels – a metaphysical view known as 'classical determinism'. If classical determinism is true, then:

> *"even though indeterminism reigns in our brains at the subatomic quantum mechanical level, our macroscopic decisions and acts are all themselves determined"* (Dennett, 2015:148)
>
> *"although there is indeterminism at the micro-level, the level of small particles, there is still determinism at the macro-level, which includes neural events and everything with which we are ordinarily familiar"* (Honderich, 1993:140)

In support of this conclusion, philosophers have provided various forms of what we will refer to as 'Las Vegas arguments' (since their objective is to ensure that what happens in the quantum realm stays in the quantum realm). Clearly, a metaphysics of classical determinism cannot just be assumed or asserted for the purposes of debate. A successful Las Vegas argument is required to motivate the assumption of classical determinism and its relevance for discussions of free will.

In the first part of this section, we argue that none of the proposed Las Vegas arguments are successful in their task. In doing so, our aim is to demonstrate that classical determinism is not something that has been conclusively proven and, as such, it is not a metaphysical constraint that naturalistic philosophers of free will *need* to abide by.

In the second part of the section, we argue further that, not only has classical determinism not been *proven* by physics, but in fact there are some very strong conceptual and empirical reasons to think that it is highly implausible as a description of the way that our macroscopic universe evolves. That is, it is not only the case that naturalistic philosophers of free will *need not* adhere to classical determinism in their work, but there are compelling reasons to think they *should not*.

### 3.1 The failure of Las Vegas arguments

Here we survey three flavours of Las Vegas argument and show why they do not hold.

**a) Classical physics *just is* deterministic.**

The simplest kind of Las Vegas argument simply rests on the (mistaken) belief that classical determinism has been empirically proven by physics. In this view, regardless of how exactly it happens, the assumption is that we just *know* that the classical world obeys its own rules, untroubled by whatever random events may be occurring at quantum levels. We just *know* that "determinism should be thought of as Nature's default mode" when it comes to macroscopic systems such as ourselves (Earman, 2008:817).

How do we know this? Because determinism has apparently been shown by, or perhaps is a well-established *result* of, the best physical theories we have for describing the macroscopic world – namely, classical physics (which includes Newtonian mechanics and Maxwell's electrodynamics). Indeed, the view that the classical world is deterministic has held sway among physicists since Newton's development of his laws of motion, which proved so effective in predicting the orbits of the planets. It might therefore seem reasonable for free will philosophers to extrapolate from Newton's success in this domain to the idea that all systems must be similarly deterministic, at least at macroscopic levels of resolution, even if their complexity makes them unpredictable in practice. As the common refrain goes, 'the (classical) universe is clearly deterministic enough for us to have gotten to the moon'.

But is it actually right to say that classical physics has *proven* that the macroscopic world of brains, bodies and behaviours evolves deterministically? Or have we perhaps been misled by an exceptionally obliging solar system, as suggested by the philosopher Elisabeth Anscombe (1971) who said that: "*[t]he high success of Newton's astronomy was in one way an intellectual disaster: it produced an illusion from which we tend still to suffer*"? In other words, might it in fact be the case that the widespread belief that classical systems are provably deterministic is one of those ideas, to use Mark Twain's phrasing, that we "know for sure, that just ain't so"?

Philosopher and historian Marij van Strien (2021) expertly addresses these questions through a historical lens, asking: "*was physics ever deterministic*?". After surveying the extent to which physics in the pre-quantum (now referred to as 'classical') period was committed to the thesis of determinism, she concludes:

> *"during the period which we now describe as classical, determinism was not so much an established result of physics, but rather an expectation, and ... during the late nineteenth and early twentieth century, it more and more took the form of a methodological principle or necessary presupposition of science, rather than an ontological claim."* (p.2)

Most importantly for our purposes, van Strien explains that the thesis of determinism, as it was originally articulated within the modern context (e.g., by Laplace), was deeply embedded within a very specific research program within classical physics – one that sought to reduce all of physics to mechanics (motion), and all of mechanics to the motion of idealised point-mass particles. Yet, even before the discovery of quantum physics at the turn of the twentieth century, these explanatory objectives had mostly been abandoned – due, among other things, to the implausibility of a solely point-mass conception of matter and a growing disbelief that statistical laws (such as the second law of thermodynamics) could ever successfully be reduced to (or derived from) the laws of mechanics (see also Drossel, 2015). Consequently, as van Strien says, *"by the end of the nineteenth century, it had become increasingly unfeasible to reduce all of physics to a basic set of equations and to establish that these equations would always have a unique solution for given initial conditions"* (p.10). In other words, the original basis for the claim that (classical) physics entails a deterministic picture of the macroscopic universe had become fundamentally implausible.

Instead, the author notes, determinism took on the role of a methodological presupposition within classical physics – an explicitly pragmatic assumption that was seen as necessary for scientists to *do* science. Many physicists in this period, including Max Planck, Ernst Mach and Henri Poincaré, publicly adopted a position of agnosticism with regard to whether the (classical) universe was *really* deterministic or not, noting only that determinism was a necessary heuristic principle for guiding their work. As Poincaré put it, "we are determinists voluntarily" (1921:347).

Of course, the fact that determinism is not, and has never been, an established *result* of classical physics does not, by itself, imply that the macroscopic world is *not* deterministic. As many have noted, proving determinism would be effectively impossible. It does, however, undercut the impression (common in philosophy) that one can safely *assume* that neurons, brains and bodies are causally insulated from quantum effects. Instead, this is something that needs to be argued for or defended. We will now turn our attention to some such attempts to defend classical determinism.

**b) Decoherence quarantines quantum indeterminacies.**

A more mechanistic Las Vegas argument appeals to notions of quantum decoherence in the so-called quantum-to-classical transition (Zurek, 1991; Schlosshauer, 2004, 2014). Decoherence is an established phenomenon in physics which describes the elimination of quantum interference effects when a quantum system interacts with its environment. The consequence is that the system starts to behave in a more classical fashion. For free will philosophers, this has sometimes been taken to imply that *"quantum effects could just…be self-canceling"* (Dennett, 2015:148) or that *"micro-indeterminacies "cancel each other out," and we get macro-level determinism"* (McKenna & Pereboom, 2016:23). Decoherence therefore seems to provide a potential argument for classical determinism that goes beyond a mere (and mistaken) appeal to an empirical *proof* of classical determinism and provides, instead, a mechanism *whereby* classical levels could become insulated from quantum indeterminacy. Such an argument is not successful, however.

The claim that decoherence resolves the quantum "measurement problem" (the fundamentally *random* emergence of actualities from a probability distribution) is not, in fact, supported (Adler, 2003; Schlosshauer, 2004, 2014; Drossel & Ellis, 2018).

> *"If we understand the "quantum measurement problem" as the question of how to reconcile the linear, deterministic evolution described by the Schrödinger equation with the occurrence of random measurement outcomes, then decoherence has not solved this problem."* (Schlosshauer, 2014:19)

On the contrary, any superpositions – i.e., the range of possible states that the system *could* be observed in – that existed *or that arise* in its microstates must still be resolved, apparently still at random. Decoherence simply means that there should not be superpositions of macrostates (like simultaneously live and dead cats, as in Schrödinger's famous thought experiment). Moreover, each resolution of quantum uncertainty will generate a new state whose future evolution is described by *a new (probabilistic) wave function*, which will in turn resolve (randomly) into some actuality, with another new wave function, and so on. If we take seriously the notion that quantum fields and systems really are the lowest level of physical reality, not just under special conditions of isolated streams of electrons or photons in the laboratory, but in more complex systems, then we must accept the constant introduction of some level of noise or randomness, not just in a once-off quantum-to-classical transition which permanently eliminates all quantum nature from the macroscopic system, but as an ongoing process.

As Sean Carroll describes: *"The rule is this: whenever we measure an observable, whatever the wave function was before the measurement, it immediately collapses onto some definite value of the quantity being observed. The new post-collapse wave function then evolves according to the Schrödinger equation, until it is observed and collapses again."*[7] (Carroll, 2024:36)

Thus, the particular mechanism of decoherence (which itself remains poorly understood and open to interpretation) simply does not support a Las Vegas-style argument for classical determinism. Quantum events do not 'cancel each other out' in a way that eliminates all presence of randomness in the deepest recesses of our being. Indeterminism is instead an ever-present feature of the fundamental matter out of which we are made.

**c) Classical determinism is statistical and emergent.**

A third kind of Las Vegas argument *accepts* that there is randomness constantly at play at quantum (or just microscopic) levels within the system, but claims that none of this actually matters for how the macroscopic variables and properties of the system will evolve over time, because all of this low-level indeterminacy simply gets averaged out or coarse-grained over, leading to deterministic dynamics at the macroscopic level. It is therefore essentially an argument akin to the mathematical law of large numbers, which states that the average of a large number of random events typically converges onto a stable, regular or 'true' value. As Philip Ball (2006:76) puts it, it "is a way of saying that pure randomness gives way to determinism if the number of random events is large". So, just as a large collection of simultaneous coin flip events reliably produces a stable and predictable outcome at the statistical level (namely, a Bernoulli distribution), so too – this version of the Las Vegas argument says – does all of the microlevel randomness within a system produce a fixed and determinate 'average' overall effect at the macrolevel.

The result is that the macroscopic system can be viewed as evolving strictly according to the (deterministic) laws of classical physics, *when observed at this macroscopic scale of*

[7] Note that while this quotation frames this in the language of an "observer", this is not an essential part of the theory; any physical interaction that resolves the probability distribution function into an actuality will suffice.

*resolution*. Philosopher Daniel Dennett gestures at this type of Las Vegas argument when he asks us to consider a robot living in what he describes as a 'deterministic world':

> "Once again we will take the world of the robot explorer, for then we can know just what we are stipulating in saying that its control system is completely deterministic: we design it to be deterministic, to be highly resistant to micro-level noise and random perturbation. (It has no built-in Geiger counters to propagate random effects; it is designed instead to damp out such effects.)" (2015:126)

For Dennett, the robot evolves deterministically *despite* the continual presence of microscopic randomness, because it is able to 'damp out such effects'. Indeed, this is how most engineering systems work. Another example might be the fact that the temperature of a gas at equilibrium (a macroscopic variable) does not change *despite* the microscopic fluctuations of its individual molecules.

The result is a kind of statistical or *emergent determinism* at the classical level, where the macroscopic dynamics are taken to be insensitive to the detailed fluctuations happening at the lowest levels and can still evolve in predictable, deterministic ways *at a certain level of resolution*. Note the difference from the idea of universal determinism, which is a bottom-up argument, the idea being that everything that happens at macroscopic levels is deterministic *because* all the low-level processes are deterministic. Here, the idea is that the highest levels can have some deterministic dynamics unto themselves, *despite* all kinds of possible noise at the lowest levels.

Is this argument successful in proving that classical determinism holds in our universe? It certainly seems to accurately describe something like the orbits of the planets, which are incredibly well described by the laws of classical mechanics, despite the fact that, say, the atmospheric conditions on any of those planets evolve non-deterministically. Those details just do not affect the planet's orbit (at least not to any extent that we would ever possibly need to care about). But can we assume that behavior observed for such simple systems can be extrapolated to more complex ones? Would free will philosophers be justified in claiming that, because planetary solar systems exhibit an emergent determinism, we can assume that brains do too? We argue in the following sections that this inference is not valid; in many classical systems (even the solar system itself), macroscopic dynamics are highly sensitive to the ways in which microscopic indeterminacies are resolved.

To sum up this section, these varieties of Las Vegas arguments fail to make the case for the insulation of the classical realm from quantum goings-on, *in toto*. Classical determinism is not, and never has been, an established *result* of physics. It is not a worldview that has been proven by classical physics or one that has been secured by the discovery of quantum decoherence. There is perhaps some evidence that a sort of emergent determinism *can* take hold in simpler classical systems; however it does not follow from this that it does so in more complex systems like the human brain and body. Hence, in conclusion, there seems no empirical requirement for philosophers of free will to accept the premise that "although there is indeterminism at the micro-level, the level of small particles, there is still determinism at the macro-level, which includes neural events" (Honderich, 1993:140). In the next section, we go one step further by arguing that, for a certain class of macroscopic system, which includes the human brain, there are also some strong, positive reasons to think that their evolution is decidedly *not* deterministic in the manner prescribed by classical determinism.

### 3.2 Direct arguments against classical determinism

Clearly, the classical determinism worldview depends on the same set of assumptions as the universal determinism worldview, modified slightly for its narrower domain of applicability. To recap, these assumptions are: (i) that (macroscopic) systems exist in precisely defined (or hypothetically *definable*) states at every given instant of time, and (ii) that the transitions between these 'states' of the (macroscopic) system are derived from (or described by) sets of differential equations which always have a unique solution for the given antecedent state (i.e., that the laws of classical physics are intrinsically and unfailingly deterministic).

The first of these assumptions could presumably be cashed out in one of two ways. First, when one speaks of the 'state' of a macroscopic system, such as a brain, one could be referring to the collective states (position, momentum, etc.) of all of the individual atomic or subatomic elements that materially constitute the system at that time. On such a view, for the 'state' of a macroscopic system to be precisely defined just *is* for each of its smallest (i.e., quantum) constituents to exist in a precisely defined state. However, as we saw in Section 2.1, such a situation is physically ruled out by the HUP.

An alternative interpretation of assumption (i) is the more commonplace view that each of the classical-level physical parameters of a macroscopic system (e.g., its position, length or centre of mass) exists in a precisely defined state, at any given time. Typically, this would mean that each of these parameters could, at least in principle, be described with infinite precision using a set of real numbers – that is, where all the decimal places are given, all at once (although cf. Kwok, 2020). We take it that this is generally what is required for a macroscopic system to exist in a precisely defined (or hypothetically definable) state. Yet, this is a situation that cannot actually hold in the physical world. Such precision is simply a mathematical idealisation. As expressed by pioneering quantum physicist Max Born (1969): "*Statements like 'a quantity x has a completely definite value' (expressed by a real number and represented by a point in the mathematical continuum) seem to me to have no physical meaning*." Likewise, Karl Popper (1950:123) said: "*infinitely precise and complete knowledge is also 'in principle' unattainable*."

To see why, consider again the argument presented in Section 2.2. for why the initial universe could not contain sufficient (i.e., infinite) information to predict every future state and event. According to this argument, such a proposition would require a finite region of space to hold an infinite amount of information – thereby violating physical law. As compellingly argued by Flavio del Santo and Nicolas Gisin (Del Santo & Gisin, 2019; Del Santo, 2021; Gisin, 2021a, 2021b), the same logic applies to any subsequent time-slice of the universe, and even more so to particular systems *within* it. Such systems occupy finite space. Yet, for the physical parameters of these systems (at both micro- and macro-scales) to exist in the precisely defined state demanded by classical determinism, it would need to be the case that these parameters can be described (at least in principle) using real numbers. Real numbers, however, typically feature no structure at all (i.e., they are fundamentally incompressible) and thus contain an infinite amount of information. It would therefore be a simple violation of physical law for the parameters of a macroscopic system to *actually* exist in a precisely defined state at any given instant of time.

These authors show, therefore, that the supposed determinism of classical physics ultimately rests on what is really just a mathematical idealisation, with no plausible basis in physical reality; namely, the idea that the real numbers which describe the physical parameters of any system are given with infinite precision, all at once. Indeed, even David Hilbert, the most ardent defender of the mathematical concept of infinity, admitted that this concept had no basis in physical reality (Ellis et al., 2018): "*The infinite is nowhere to be found in reality, no matter what experiences, observations, and knowledge are appealed to*." Or, as Nicholas Gisin (2021b) puts it, "*real numbers are not really real".*

The upshot is there seems to be no way, in principle, that the state of a classical system can ever be defined (or, indeed, *exist*) with sufficient precision to meet the job description of determinism. The so-called 'initial conditions' of the system are always going to be somewhat fuzzy, vague and undefined – they exist in well-defined states up to a certain degree of resolution, but at higher levels of precision their state is truly indeterminate (Mariani & Torrengo, 2021). That is, after a certain number of decimal places, the system's state will just be undefined or indefinite (Ben-Yami, 2020).

Under this view, classical determinism becomes fundamentally impossible as an ontological claim – an artefact of our mathematical idealisations, rather than a feature of the world. As physicist Barbara Drossel (2015) explains:

> *"In classical mechanics, the state of a system can be represented by a point in phase space. The phase space of a system of N particles has 6N dimensions, which represent the positions and momenta of all particles. Starting from an initial state, Newton's laws, in the form of Hamilton's equations, prescribe the future evolution of the system. If the state of the system is represented by a point in phase space, its time evolution is represented by a trajectory in phase space. However, this idea of a deterministic time evolution represented by a trajectory in phase space can only be upheld within the framework of classical mechanics if a point in phase space has infinite precision. If the state of a system had only a finite precision, its future time evolution would no more be fixed by the initial state, combined with Hamilton's equations. Instead, many different future time evolutions would be compatible with the initial state."* (p.2)

Moreover, if *"many different future time evolutions would be compatible with the* [same] *initial state*" then the second key assumption of classical determinism also seems under threat. This assumption required that the time evolution of a macroscopic system proceeds *necessarily* from one timepoint to the next – governed by some sort of deterministic causal laws. However, as Drossel explains, such a worldview would be highly infeasible if, as we have argued above, the initial conditions of the system cannot be specified with infinite precision. For such systems, "the state [of the system always] has a fundamental indeterminacy that leads to an indeterministic dynamics" (Del Santo, 2021:67). This is most noticeably true for systems with "chaotic" dynamics, such as the brain, as we explore in the next section.

### 3.3 Chaotic systems

The importance of this fundamental imprecision or indefiniteness for the problem of free will becomes especially acute when we consider that biological systems are inherently chaotic, dominated by non-linearity and a sensitivity to initial conditions (Deco & Kringelbach, 2020; Terada & Toyoizumi, 2024; Deco et al., 2025). This sort of sensitivity was famously discovered by Lorenz in running computer simulations of a simple weather system (Lorenz, 1963, 1969). He found that re-running a simulation from a given point, but with the initial parameters truncated after fewer decimal points than in the first run, produced a trajectory of the system that initially followed that of the first run closely, but that then began to diverge, such that, after some period of simulated time, there was no longer any correspondence at all between the two trajectories. That is, the evolution of the system as a whole was exquisitely sensitive to tiny changes in the starting parameters, to the point that, beyond a certain time horizon, two systems that differ only fractionally in their initial conditions "*will evolve into two states differing as greatly as randomly chosen states of the system*" (Lorenz 1969:289; see also Palmer et al., 2014). Such systems became known as chaotic systems.

It is often claimed that, despite being unpredictable in practice, the evolution of these chaotic systems is still nevertheless deterministic. That is, their future *is* actually fixed by their initial conditions – it is just that *we* cannot know what those are in full detail (i.e., the observed indeterminacy is merely epistemic in origin). This kind of deterministic situation may in fact hold for computer *simulations* of such systems, where any given set of initial conditions will reliably produce the same results over multiple runs (to the level of precision being simulated). However, in the real physical world, if the parameters describing a system do not in fact *exist* in the world with infinite precision, as we have argued above, then such *ontological* imprecision will necessarily lead, in a system with chaotic dynamics, to a *genuine under-determination* of future trajectories by the present state of the system (Del Santo & Gisin, 2019; Ben-Yami, 2020; Gisin, 2021b). In such cases, the evolution of the system will be exquisitely sensitive to details (e.g., the value of the one hundredth digit describing some macroscopic parameter) which are not just unknown to us, they are unknown to the universe itself – they are ontologically fuzzy and undefined states. In one striking example of this, "gargantuan" simulations of a three-body system (three gravitationally interacting black holes) have shown a sensitivity to the details of initial conditions that is *below the Planck* *length*, the smallest possible subdivision of space. Such a physical system literally could not exist in a sufficiently defined initial state for its chaotic evolution to be 'actually' deterministic (Boekholt et al., 2020).

In fact, this is even true in the solar system itself. Henri Poincaré famously recognised the "three-body problem", which results from the non-linear interactions between more than two gravitating bodies. The evolution of such systems (which includes our solar system) is, in general, highly sensitive to even slight variations in initial conditions. Because the mass of our sun dominates the gravitational forces in the solar system, the inherent indeterminacy in the evolution of these planetary orbits may take hundreds of millions of years to manifest, but it is still a real feature of their dynamics (Laskar, 1990, 2013; Boué et al., 2012). Newton himself recognised this, highlighting the import of "inconsiderable Irregularities…, which may have risen from the mutual Actions of Comets and Planets upon one another, and which will be apt to increase, till this System wants a Reformation." (Newton, 1952:402).

Indeed, Zhihong Xia showed that a mathematical model of a system of five point masses, can, under highly specific starting conditions, lead to singularities in finite time even in the absence of collisions (Xia, 1992)[8]. That is, Newton's equations can under some circumstances inherently fail to produce a well-defined unique evolution for all times – there simply is no solution beyond some finite time. The limitation here is not merely practical unpredictability, but a breakdown of the theory's ability to specify the system's evolution at all. This result undermines the assumption that determinism for all times follows automatically from classical laws of motion.

Thus, it is quite possible for a physical system to be both chaotic *and* fundamentally indeterministic, and indeed, the systems we are interested in – living organisms and the local world they inhabit – clearly are both. This means that the evolution of their physical states through time is genuinely under-determined not just at microscopic, but also *at macroscopic levels*. It is simply an intrinsic and irreducible feature of chaotic systems in our universe that – absent any top-down constraints emerging from selection for macroscopic functionalities – their future is under-determined by their current physical state. (Note that the emergence of such constraints is precisely what characterises life and the exercise of choice and control, as we discuss in detail in a companion paper (Potter & Mitchell, 2025a)).

### 3.4 Determinism does not hold at any level

This survey of principles of quantum and classical physics shows that any strict version of the thesis of determinism – either universal *or* classical – rests on assumptions and mathematical idealisations that simply do not seem to hold in physical reality. As John Earman put it: "*determinism succeeds only with a little – or a lot – of help from its friends*" (Earman, 2004:12) – where claims of determinism often amount to little more than "*making a postulate of wishful thinking*" (p.4).

[8] We thank a reviewer for alerting us to this paper.

Similarly, Barbara Drossel (2023) argues: "*even though the supposedly fundamental theories are deterministic and time-reversible, these theories are in practice supplemented by irreversible and stochastic features. There is no reason to assume that irreversibility and stochasticity are only apparent. Such a claim is based on ideology and not on evidence*." (pp.13-14)

Contrary to convention in the philosophy of free will then, physics has not proven that complete determinism holds *at any level*. Indeed, the evidence from physics itself strongly suggests that strict determinism at both the quantum and the classical level *must* be false. The current state of any system plus the laws of physics do not, in fact, specify a single future for all subsequent time points in infinite detail – instead, they under-determine it, meaning genuine alternate possibilities exist and the future is radically open, from the perspective of physics. For some kinds of systems – like the orbits of the planets – the behaviour of the system at macroscopic levels will be *effectively* deterministic (such that, for example, the position of the centre of mass of the Earth within the solar system is highly predictable many centuries in advance to some finite degree of precision). But for chaotic systems – like atmospheric systems or nervous systems, or even the solar system considered over longer timeframes – their future states are genuinely under-determined by the laws of physics themselves. There is thus no good reason to accept classical determinism as our starting premise or focal point for discussions of free will, and many good reasons not to.

## 4 Determinism-plus-randomness

In subsection 4.1, we examine the influence of the determinism-plus-randomness worldview within the philosophical literature on free will. In subsection 4.2, we present arguments from physics against determinism-plus-randomness. And in subsection 4.3, we propose an alternative model of indeterminism, which we call pervasive indefiniteness.

### 4.1 The premise of determinism-plus-randomness in the free will debate

Where does this leave us? We have so far argued, in contrast to the literature's current conceptual terrain, and crucially *for reasons that are independent of any considerations about free will itself,* that philosophers of free will can and should abandon their interest in "the thesis that there is at any instant exactly one physically possible future" (van Inwagen, 1983:3). If our arguments are right, then such determinism is simply not an accurate description of our macroscopic universe or of the universe as a whole. It is just not the case that everything in the universe evolves *necessarily* from one timepoint to the next (*contra* universal determinism), or that there exists indeterminacy at the lowest levels of reality but not at the higher, macroscopic levels (*contra* classical determinism).

For many, this conclusion may not be especially newsworthy (though we hope the details of the arguments presented above will help permanently exorcise Laplace's demon). That is because, as noted in the introduction, many philosophers of free will are not officially committed to the truth of determinism – despite the structure of the theoretical landscape being as it is. Many acknowledge that our neural decision-making processes are not fully deterministic, and some (namely, libertarian philosophers of free will) even actively embrace indeterminism as an essential ingredient in their theories of free will.

However, in doing so, it seems to us that these philosophers typically rely on a picture of indeterminism that, to a surprisingly large extent, continues to posit the reality of (classical) determinism – a view which we have called determinism-plus-randomness. On this view, randomness or indeterminacy is framed as something that gets *added* to an otherwise deterministic universe or an otherwise deterministic decision-making process. Consider, for example, libertarian philosopher David Palmer's definition of the libertarian view of free will as one which *requires* "our actions to be breaks in the deterministic causal chain" (Palmer, 2014:4), or his colleague Robert Kane's claim that "indeterminism does not have to be involved in all acts done "of our own free wills"" (2019:147). What both of these conflicting statements share, we suggest, is the presupposition that determinism *remains* "Nature's default mode" (Earman, 2008:817), at least for macroscopic systems such as ourselves. For Palmer, determinism holds *except* in cases of freely willed action. For Kane, determinism holds *even* in the case of some (but not all) freely willed actions. The determinism-plus-randomness worldview therefore appears to differ from classical determinism with respect to only one additional claim: that deterministic goings-on *can sometimes* be disrupted, broken, or diverted by the occasional insertion of randomness into the relevant causal chains. (Hence why we categorised this worldview in the introduction as a version of 'near-determinism'.)

The visual metaphor that is often used to illustrate the determinism-plus-randomness concept of indeterminism is, after Borges, what we call a 'Garden of Forking Paths' (see Figure 2)[9]. In this, "the single line going back into the past is just that: a single line indicating "same past"; while the multiple lines going into the future represent "different possible futures"" (Kane, 2007:24; see also van Inwagen, 1990; Law, 2023). The points of bifurcation therefore represent occasions where randomness 'percolates up' to have an effect at the macroscopic scale, creating diverging paths along which determinism holds until the next random bifurcation is reached.

While presumably not every libertarian endorses this perspective, a commitment to such a worldview is certainly evident in the ongoing debates among event-causal libertarian philosophers (and their critics) over *where* in the causal chain leading to an action indeterminism needs to occur in order for an action to be considered free (in the libertarian sense). Or, as Laura Ekstrom (1999:85) puts it: "Where precisely are the *gaps* that must exist in nature in the chain of deterministic causal links between events in order to allow for human free will?". This 'location question' is often taken to be one of the main problems that any viable libertarian account of free will must address (Franklin, 2013, 2018; Kane, 2016).

[9] Though we note that in Borges' story, all paths happened, while here we take them to be exclusive alternate possibilities.

And, indeed, one of the most common ways to group the different (event-causal) libertarian theories of free will is according to where exactly they *locate* the indeterminism (Franklin, 2018; Clarke et al., 2021), with so-called "deliberative" theories locating the relevant indeterministic bifurcation or 'branching point' in the early stages of a decision-making process (Ekstrom, 1999; Mele, 2006; Dennett, 2015) and "centred" accounts placing it at the moment of the decision or action itself (Kane, 1996, 2007, 2019; Balaguer, 2010; Franklin, 2018; Ekstrom, 2019).

Implicit in all of these disputes is an acceptance that, wherever the indeterminism is *not* located, determinism holds true. And wherever it *is* located, it takes the form of a random or probabilistic *intrusion* that disrupts the otherwise deterministic goings-on. But is this the right way to conceptualise indeterminism? If the arguments presented so far in this paper are right, then we suggest it is not.

### 4.2 Arguments against determinism-plus-randomness

It is certainly true that isolated, particular quantum events can have influences on macroscopic systems. Obvious examples include the emission of a photon by an excited atom or the radioactive decay of an atomic nucleus, which are inherently random in both the time and direction of emission (Ginzburg & Syrovatskii, 2013). One need not conjure contrived scenarios like Schrödinger's cat to see how such events could have an impact at macroscopic levels. Cosmic rays from distant systems can, for example, alter electronic states in digital computers, leading to so-called soft errors in computer memory (Ziegler & Lanford, 1979; Baumann, 2005). Cosmic rays can also cause damage to DNA, leading to macroscopic effects such as cancer (Percival 1991; Atri & Melott 2014), or even contributing to the course of evolution of species.

However, these scenarios refer to outcomes of singular random quantum events impinging on a system from outside. A more pressing question is whether individual quantum events occurring *within a system* can 'percolate up' to affect macroscopic processes, in the manner envisaged by determinism-plus-randomness. In the brain in particular, we can ask if an individual ion in some neuron 'zigs' instead of 'zags', could that alter the firing of that neuron at some crucial moment, in a way that could ultimately make the difference between someone deciding to do A rather than B? There seems to be no reason to take that picture as realistic. First, while the processes of neural transmission and firing are noisy (Faisal et al., 2005; Deco et al., 2009; Rusakov et al., 2020), there is little evidence that they can generally be swayed by *single events* at the quantum level. And, second, while decision-making processes are similarly noisy (Glimcher, 2005; Sanborn et al., 2024), there also is no evidence – in mammals at least – that they are generically sensitive to single firings of single neurons. On the contrary, neural systems are often designed to be largely insensitive to the precise details of individual neuronal firings (Potter & Mitchell, 2025b). Even Epicurus, who proposed that atoms must occasionally "swerve" to "loosen the treaties of fate" and thus open the possibility of free will, did not envisage the kind of <one swerve-one action> relationship implied by determinism-plus-randomness (Sedley, 1983).

Instead, the empricial evidence surveyed above suggests that, rather than fixating on isolated random events at quantum levels, we ought to understand indeterminism in terms of the much more ubiquitous probabilistic nature of microscopic goings-on, which *collectively* entail an under-determination of the macroscopic evolution of the system by its current states. In a system where definite things are nevertheless happening, that fundamental indefiniteness necessarily gets resolved through interactions, apparently genuinely at random. The question then is whether and how this very general and pervasive process manifests at macroscopic levels.

As discussed in Section 3.3, the answer lies in the fact that many systems are chaotic, meaning that the evolution of their macroscopic states are inherently sensitive to the ways in which these low-level indeterminacies get resolved, *by default*. This is especially true for systems characterised by turbulence and 'spontaneous stochasticity', where it remains unknown whether the classical equations of fluid dynamics have unique solutions for three-dimensional systems (Eyink & Drivas, 2015; Neyrinck et al., 2022; Bandak et al., 2024). In biological systems in particular, the constant broiling of thermal fluctuations adds to this stochasticity. Large biological molecules in water (essentially their context in a cell) undergo ~$10^{14}$ collisions per second (Graham, 2023:25). Given the uncertainty relation between position and momentum discussed above, these interactions will be fundamentally (not just epistemically) indeterministic. For turbulent systems, this means that their evolution at macroscopic scales will be genuinely and inherently indeterministic (Bandak et al., 2024), as we saw for other kinds of chaotic systems above.

Macroscopic brain dynamics can also formally be described as 'turbulent' in direct analogy to such systems (Deco et al., 2009; Deco & Kringelbach, 2020; Deco et al., 2025), with neural networks often being poised at 'criticality', optimising stability for maintaining ongoing processes and flexibility for rapid adaptation (Hesse & Gross, 2014; Terada & Toyoizumi, 2024). Thus, the kinds of systems most relevant to discussions of decision-making and free will are chracterised by a *generic* noisiness of underlying processes (Deco et al., 2009; Tsimring, 2014), and a consequent openness of macroscopic evolution, rather than a sensitivity to *isolated* quantum events.

### 4.3 Pervasive indefiniteness

The empirical arguments presented above thus support a far more radical concept of indeterminism that libertarians can make use of.[10] According to this conception, classical scales of reality are *themselves* indeterministic. First, in the sense that every classical system exists in a state that is, to some degree, indefinite: the states and properties of the system are pervasively 'fuzzy' insofar as that they lack complete ontological precision. And second, for non-linear macroscopic systems such as the brain specifically, this pervasive indefiniteness means that the way that the physical states of these systems will evolve over time is then *by default* under-determined by the low-level details of their current state. This means that 'Nature's default mode' for an agent's neural decision-making processes is simply *not* deterministic, as is supposed by a determinism-plus-randomness worldview.

[10] Moreover, as argued in Potter & Mitchell (2025a), we contend that libertarians *should* make use of this alternative concept of indeterminism in order to fully overcome the infamous Luck Objection.

Under this new concept of indeterminism, it is therefore not appropriate to assume that decision-making processes would have "exactly one physically possible future", if they were somehow to unfold uninterrupted by isolatable intrusions of randomness. This is just an idealisation. On the contrary, for processes such as these, the 'default mode' is an indeterministic time evolution – and, as such, our baseline expectation should be that, given the system's current state, at any given timepoint, there will always be multiple physically possible ways it could evolve. To use Müller et al., (2019)'s terminology, our argument is that agents like us *inevitably* have a wide range of "real possibilities" available to us, where "what is *really possible* in a given situation is what can temporally evolve from that situation against the background of what the world is like" (p.3).

What this more radical conception of indeterminism means for free will philosophy is subtle but crucial. First, it means that mechanisms for the amplification of quantum effects need not be identified and cited in order to get off the ground the idea that the future of a macroscopic system is metaphysically open. This sort of openness comes for free in virtue of being a non-linear system in a universe where one's physical parameters necessarily contains *some* degree of ontological imprecision. Second, it means that indeterminacy should not be viewed as something that gets *added* intermittently to the evolution of macroscopic systems – it is not the result of an extra, *positive* sort of randomness that somehow intrudes into our universe to trouble the otherwise fully deterministic go of things, occasionally forcing upon us unwilled actions. It is, rather, a *negative* – a constant under-determination of the future by the low-level details of present states and the laws of physics, entailed by the unavoidable and pervasive indefiniteness (and thus openness) that characterises the present state of the system itself (Mariani & Torrengo, 2021).

Note that this does not imply that *just anything could happen* at any moment. Rather, at any moment, there is a possibility space that is not restricted to *a single option*, defined with infinite precision. Each actual outcome generates a new possibility space. The fuzziness of the possibility space thus expands as we look further into the future, until we reach a point where some future states are simply currently undefined – they have no logical truth-value. There may, for example, be a definable probability that it will rain tomorrow in Dublin, but there is no definable probability that it will rain a year from today. It is still not the case that *just anything* could happen – it's that determining *what will happen* within that general possibility space requires the system to actually traverse some actualized trajectory, defined by all kinds of stochastic contingencies (i.e., the repeated resolution of probability distributions to actual states of affairs).

One notable upshot of conceptualising indeterminism in this way is that the 'Garden of Forking Paths' is no longer the right metaphor for visualising the future of an indeterministic system. Instead, as Mitchell (2023:161) describes, *"[i]f we really could glimpse the future, we would see a world out of focus. Not separate paths already neatly laid out, waiting to be chosen—just a fuzzy, jittery picture that gets fuzzier and jitterier the further into the future you look*".

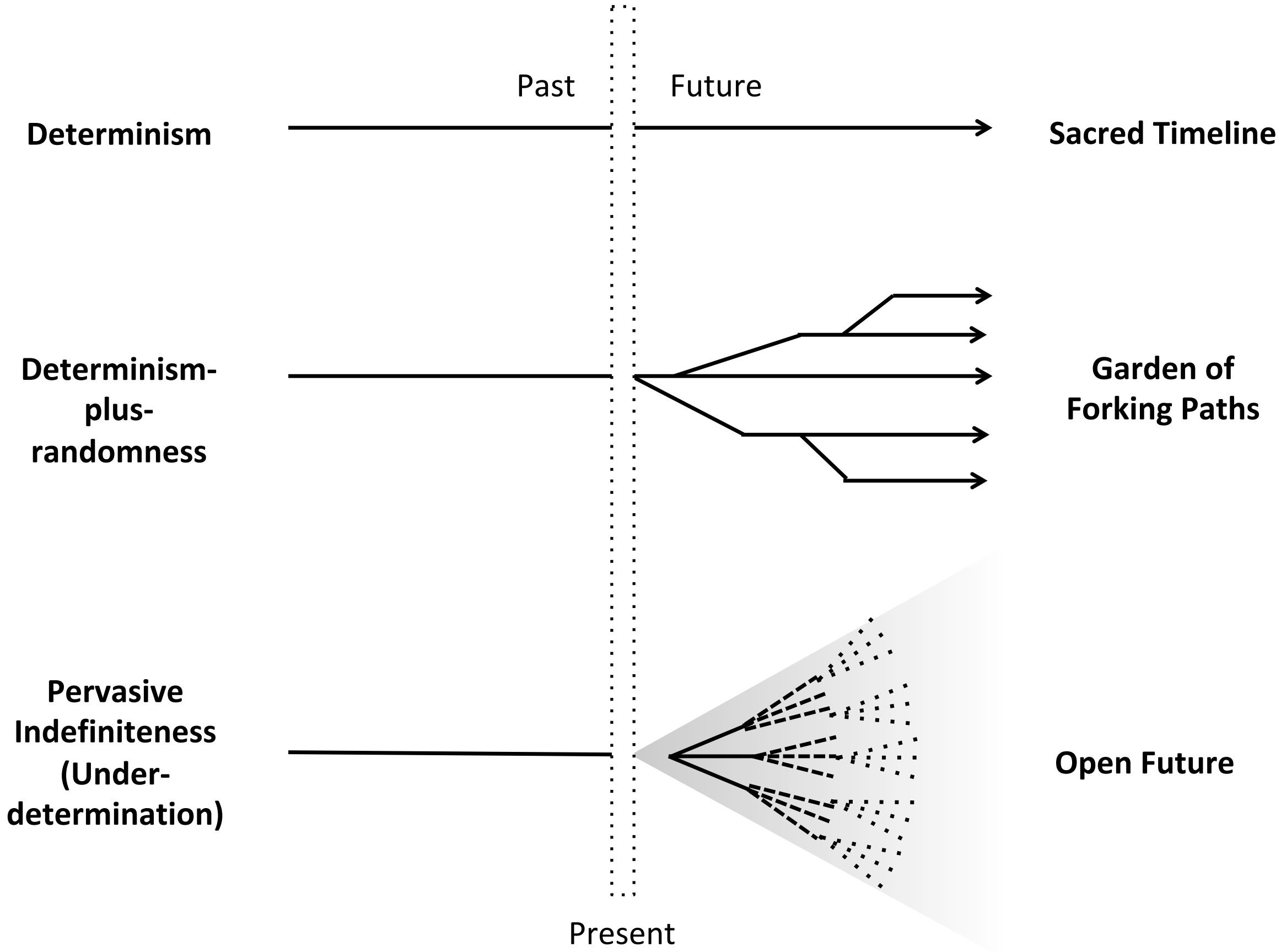

**Figure 2.** Top: Universal (Laplacean) determinism posits a single, unbranching timeline, with no *real possibilities*. Middle: The standard way to conceptualise indeterminism ("determinism-plus-randomness") sees the future as a series of forking paths, where alternate possibilities in a sense are pre-existing or pre-statable, and which branch is taken is determined by the intrusion of randomness at specific moments into an otherwise deterministic evolution. Bottom: An alternative way to conceptualise indeterminism ("pervasive indefiniteness") which sees the future as radically open; with the near-future tightly constrained by current states, but the far-future highly under-determined beyond some time horizon. The present is then defined as the time during which events *occur*, resolving the open possibilities into actual happenings, which then become the fixed past (Smolin & del Verde, 2021; Mitchell, 2023). In this view, in the present "indeterminacy gets resolved as particles catch each other mid-jitter and interact to form some new state, the components of which start jittering all over again" (Mitchell, 2023:163).

## 5. Implications for the free will debate

The arguments above have important implications for the framing of the free will debate. We contend that they render many of the historical arguments and dividing lines in this field moot. First, if determinism is false, then the existence question – *do we have free will?* – simply ***does not entail or rest on*** the traditional compatibility question – *is free will compatible with determinism?* – as it has traditionally been assumed to. It thus makes little sense to continue carving up the theoretical terrain primarily in terms of this question, as is currently the case where "the basic divide among philosophers is between compatibilism and incompatibilism" (Palmer, 2014:4).

Put another way, if our arguments above are correct, then there is simply no problem posed by causal determinism to the existence of free will, because the former does not hold. There is thus no need for philosophers to continue focusing so extensively on mounting attacks and defences of either free will itself or of moral responsibility against this non-existent threat. Instead, compatibilist philosophers in particular may be freed up to turn their attention toward a question that often gets gestured at in the literature, but is strikingly rarely given a concerted treatment: *are compatibilist accounts of free will compatible with indeterminism?* (but cf. Fischer, 2014; Sartorio, 2021; see also Mackie, 2018 for an in-depth discussion of the problem at hand).

Moreover, if we do not start with the defunct premise of determinism, then we no longer have to wonder: where does the 'freedom' required for free will come from? In an indeterministic universe, freedom (in the broad sense of *real possibilities* or *leeway*) comes for free. The more pressing question for a naturalistic philosophy of free will is instead: where does the control come from? How can agency emerge in the presence of such indeterminacy?

Crucially, answering this question depends on one's understanding of the nature of the indeterminacy itself – one's *concept* of indeterminism. And, thus, the second major implication for the free will debate that emerges from the evidence and arguments surveyed above is the introduction of a new way of conceptualising indeterminism. Specifically, we have argued that randomness is not something 'added' to the world, something 'extra' that 'gives' you freedom. It is the noisy backdrop against which everything the organism does is set; a pervasive and irreducible feature of chaotic systems like us, and the challenge we are constantly facing. In a companion paper to this one (Potter & Mitchell, 2025a), we show how this conceptual shift can help to flip the script on the Luck Objection that has traditionally plagued libertarian accounts of free will (Mele, 2006; Levy, 2011). In a world where the future is radically open, and a myriad possible things inevitably *could happen*, the organism must exert control to make happen *what it wants to happen*.

When seen through this lens, it becomes clear that the fundamental indeterminacy of low-level goings-on is the very thing that allows for macroscopic control – and thus organismal agency and free will – to emerge. Indeterminacy and randomness, *on their own*, are not sufficient for (or even overly conducive to) an agent's acting freely. Instead, what is key about the pervasive indefiniteness concept of indeterminism, as was recognised by Epicurus over two thousand years ago (Sedley, 1983), is that it means the low-level laws of physics are not “causally comprehensive” (contra Carroll, 2021). This then enables the way systems are *organised*, at macroscopic scales, to also have some causal power in determining how things go – not by changing the underlying laws of physics, but simply by adding higher-order constraints (Tse, 2013; Ellis, 2016; Juarrero, 2023; Potter & Mitchell, 2025b). It is precisely this kind of “causal slack” that allows self-governing systems to emerge – ones that are capable of directing their own evolution at macroscopic scales, without having to control every microscopic element (Ellis, 2016; Ismael 2016; Mitchell, 2023). In Potter & Mitchell (2025a) we show how such control can emerge and be exercised both diachronically, in the development of guiding reasons, and synchronically, in the execution of real-time decisions.

## 6. Possible Objections

We first look at some possible objections from physics to our arguments against determinism (§6.1-§6.3), and then at a possible philosophical objection (§6.4).

### *6.1. Unitarity and the conservation of information*.

A common objection to the idea of fundamental indeterminacy in the evolution of quantum systems is that the Schrödinger equation is “unitary”, meaning that information should be conserved as the equation follows the evolving system through time (Chiribella & Scandolo, 2015). This unitarity is taken to be such an essential feature of quantum theory that it is simply an unquestionable article of truth. And of course it does actually hold, right up to the moment when something actually happens, at which point the non-unitary collapse of the wave function occurs (Kastner, 2017), which absolutely does not conserve information (Drossel & Ellis, 2018).

Note that this view does not depend on an ontologically real interpretation of the wave function. We can treat it simply as a convenient mathematical object. The important point is that its characteristic unitarity does not apply to actual happenings. A similar feature emerges in the “division events” in the stochastic-quantum correspondence theory of Barandes, which interrupt the ongoing stochastic evolution of a system (Barandes, 2023; 2025). In either formulation, information is both gained and lost in such ‘division’ or ‘observation’ events. Gained in that the uncertainty of what will arise from a given probability distribution is resolved. And lost in the sense that that distribution cannot then be recovered from the resulting state of the system.

### *6.2 Time-reversibility*

A related objection is that the equations of physics are time-reversible, which is again taken to mean that you cannot genuinely lose or create information as you go along. While it is true that *the equations* are time-invariant and symmetric, that does not mean the world is. As soon as people want to actually calculate or predict the behavior of real physical systems, they have to supplement those equations with irreversible and stochastic features (Ellis & Drossel, 2020). The evolution of some systems clearly involves time asymmetric events, like division events or the collapse of the wave function.

### *6.3. The Block Universe*

Einstein pictured the universe as a four-dimensional "block" of spacetime, with essentially no arrow of time (and no explanation for why it is always "now"). This supports a Laplacean position, where all of existence is simply given at once, thus reinforcing deterministic intuitions. There is, however, no need to accept this metaphysical interpretation, especially as it fails to explain why we occupy a particular time-slice that we call "the present". Many alternatives exist, which do entail genuine arrows of time (Layzer, 1975). These include, for example, the Evolving Block Universe (Ellis & Drossel, 2020), which expands both in space *and in time*. The time reversal invariance of the underlying equations is broken by the existence of a global arrow of time, established through the expansion of the universe through a hot early state to its present state. It is a view that recognises the present as the time in which the indefinite future becomes definite through physical interactions (or "events") as in the theory of Smolin and Verde (2021).

### *6.4 Causal Determinism*

Some philosophers may object that we have been confusing the thesis of physical (pre-) determinism with the more general thesis of causal determinism. According to the former, the universe is such that 'there is at any instant exactly one physically possible future'. According to the latter, as described by Leibniz's somewhat tautological principle of sufficient reason: "*There is nothing without a reason, or no effect without a cause*". (The tautology lies in assuming that every event is necessarily an *effect*, which then obviously implies the existence of a cause). An objector might contend that it is this latter notion of causal determinism that is at the heart of the free will debate and so our arguments have missed their mark.

Our response depends on what exactly the thesis of causal determinism is taken to imply. If it is taken to mean that there are no effects (events) that are not *necessitated* to occur by some antecedent cause (or set of causes), then we would argue that that just *is* a re-statement of the thesis of physical (pre-)determinism (D'Ariano, 2018), and we hope to have shown why this is not an accurate description of our universe. If, on the other hand, the claims of causal determinism are meant to refer merely to the existence of cause-and-effect relationships in our universe (including probabilistic ones), then we would agree that our arguments do not speak against that. We are not claiming that human beings or other organisms are somehow exempt from such causalism, generally. We instead follow Anscombe (1971) and others in rejecting the common underlying suggestion that "being caused implies being fixed in advance" (Runyan, 2024:112). Agents are indeed part of the causal nexus of the universe, but a part that can itself act as a cause.

**Acknowledgements:** We are grateful to the anonymous reviewers for their careful engagement with the paper and for a number of valuable and constructive critiques. We would also like to thank Keith Farnsworth, Sergei Levin, and Kenneth Silver, as well as audiences at the *Free Will: New Perspectives from Philosophy, Biology and Neuroscience* conference (Vienna, Austria) and the *Annual Meeting of the Society for Philosophy & Neuroscience* (St Louis, USA), for helpful comments and suggestions on earlier versions of this work. HP was supported by a Provost's Fund grant (1481.9050961) from Trinity College Dublin.